\documentclass[twocolumn,aps]{revtex4}
\usepackage{amssymb}
\usepackage{amsmath}
\usepackage{graphicx}
\usepackage{lscape}
\usepackage{booktabs}
\begin{document}
\title{Properties of light (anti)nuclei and (anti) hypertriton production in Pb-Pb
collisions at $\sqrt{s_{\rm{NN}}}$ = 2.76 TeV}
\author{Zhilei She$^{1}$,  Gang Chen$^{1,2}$\footnote{Corresponding Author:
chengang1@cug.edu.cn}, Hongge Xu$^{1}$, Tingting Zeng$^{1}$, Dikai
Li$^{1}$}
\address{
 $^1$School of Mathematics and Physics, China University of Geosciences, Wuhan
430074, China.\\$^2$Key Laboratory of Quark and Lepton Physics
(MOE), Central China Normal University, Wuhan 430079,China}

\begin{abstract}
We investigate the properties of light (anti)nuclei and
(anti)hypertriton production in Pb-Pb collisions at
$\sqrt{s_{\rm{NN}}}=2.76$ TeV, based on the parton and hadron
cascade and dynamically constrained phase-space coalescence
({\footnotesize PACIAE+DCPC}) model. We found that the yields of
light (anti)nuclei and (anti)hypertriton strongly depend on the
centrality, i.e., their yields decrease rapidly with the increase of
centrality bins; but their yield ratios are independent of
centrality. The results of our theoretical model are well consistent
with {\footnotesize ALICE} data. Furthermore, we found that the
integrated yields of (anti)nuclei per participant nucleon increase
from peripheral to central collisions and more rapidly with
increasing mass number. The transverse momentum distributions of the
$\bar{d},d$, $\overline{_{\overline\Lambda}^3 H}$, ${_{\Lambda}^3
H}$, ${\overline{^3 He}}$ and $^3{{He}}$ are also discussed in the
0-10\% most central Pb-Pb collisions. The coalescence parameters
$B_A$ of light (anti)nuclei and (anti)hypernuclei are analyzed.

\end{abstract}

\maketitle

\section{Introduction}
Antimatter production has received considerable attention in
particle and nuclear physics, astrophysics, cosmology and other
fields of modern physics, since Dirac predicted the existence of
negative energy states(i.e., antimatter) of electrons in 1928.
%%, such as the analysis of asymmetry
%%violence of the present observable universe with no significant
%%amount of antimatter and the exploration of dark matter.
%%Four years later, Anderson observed positron with a cloud chamber in
%%cosmic rays.
The antiprotons~\cite{1955} were discovered in 1955, and
antineutrons~\cite{1956} were discovered in 1956, followed by a
series of antimatter nuclei including anti-deuterons, anti-triton,
the anti-helium-3 in scientific
experiments~\cite{1965,prl14,1974,1970}. The hot and dense matter
forms in ultra-relativistic heavy-ion collisions, which contains
roughly equal numbers of quarks and anti-quarks~\cite{2005}, is
similar to the fireball environment in
the initial stages of the Big Bang.%%~\cite{1927}.However, the
%%progress of antimatter decoupling quickly from matter, and avoiding
%%annihilation is allowed in the relatively short-lived expansion of
%%collision experiment~\cite{2011}.
Thus, the ultra-relativistic eavy-ion collisions provide a good
experimental condition~\cite{2011} to study the production of light
(anti)nuclei and the evolution of the early universe.

%In recent years, with the development of accelerator and detector,
%the light (anti)nuclei on the level of atomic nucleus in the
%collision experiments was detected successfully, leading to a new
%study.
In 1995, the first production of anti-hydrogen atoms was
detected~\cite{1996}, and then which was trapped successfully with a
confinement time of 172~ms by the ALPHA~\cite{2010} at the European
Organization for Nuclear Research (CERN). Moreover, the ALICE
Collaboration has also published its preliminary $\overline d$ yield
of $\sim6\times 10^{-5}$ measured in the proton-proton collisions at
$\sqrt s=7$~TeV~\cite{jpgnpp}. At BNL(the Brookhaven National
Laboratory), the STAR Collaboration has reported their measurements
of $\overline{_{\overline\Lambda}^3H}$~\cite{sci328} and
$\overline{^4He}$~\cite{2011}in Au-Au collisions at the top RHIC
energy, respectively.

In theory, firstly the nucleons and hyperons are usually calculated
with some selected models, such as the transport model. Then, the
light nuclei (anti-nuclei) are studied by the reasonable hadron
final-state coalescence
models~\cite{pr1291,pr1292,prl1976,prc77,plb79}, such as the
phase-space coalescence model~\cite{prc55,prc73,plb684} and/or the
statistical model~\cite{prc81,plb697}, etc. For example, the
production of light nuclei (hyper-nuclei) in the Au-Au and Pb-Pb
collisions at relativistic energies was described theoretically by
the coalescence + blast-wave method~\cite{prc85} and the UrQMD-hydro
hybrid model + thermal model~\cite{plb714}, respectively.

Recently, we have proposed an approach studying the light nuclei
(anti-nuclei) production in the relativistic heave ion collisions by
the dynamically constrained phase-space coalescence
model({\footnotesize DCPC})~\cite{prc2012}, which is based on the
final state hadronic generated by a parton and hadron cascade model
{\footnotesize PACIAE}~\cite{cpc183}. Using this method, the light
nuclei (anti-nuclei) yields, transverse momentum distribution, and
rapidity distribution in non-single diffractive proton-proton
collisions at $\sqrt{s }=7$~TeV~\cite{prc2012} are predicted. The
light nuclei (anti-nuclei) and hypernuclei (anti-hypernuclei)
productions~\cite{prc86}, their centrality
dependence~\cite{prc88chen} and their mass number scaling
property~\cite{chen2014} in the Au-Au collisions at
$\sqrt{s_{\rm{NN}}}=200$~GeV are also investigated. Moreover,The
energy dependence of the ratio for antiparticle to particle is
studied in the high energy proton-proton collisions~\cite{wang2014}.
In this paper, we will use this method to investigate the light
nuclei (anti-nuclei) and hypernuclei (anti-hypernuclei) productions
and the properties in the Pb-Pb collisions at
$\sqrt{s_{\rm{NN}}}=2.76$ TeV.

The paper is organized as follows: In Sec.~II, we briefly introduce
the {\footnotesize PACIAE} and {\footnotesize DCPC} model. In
Sec.~III, our calculated results of light nuclei (anti-nuclei) and
hypertriton (anti-hypertriton) are presented, such as their yields, ratios, the transverse
momentum distributions, and are compared with the
ALICE data. In Sec.~IV, a short summary is given.

\section {MODELS}
The {\footnotesize PACIAE} model~\cite{cpc183} is based on
{\footnotesize PYTHIA}6.4~\cite{jhep} and is designed mainly for
nucleus-nucleus collisions. In the {\footnotesize PACIAE} model, the
process is decomposed into four steps. Firstly, the nucleus-nucleus
collision is decomposed into the nucleon-nucleon($NN$) collisions
according to the collision geometry and $NN$ total cross section.
Each $NN$ collision is described by the {\footnotesize PYTHIA} model
with the string fragmentation switches off and the diquarks
(antidiquarks) randomly breaks into quarks(anti-quarks). So the
consequence of a $NN$ collision is a partonic initial state composed
of quarks, anti-quarks, and gluons. Provided all $NN$ collisions are
exhausted, one obtains a partonic initial state for a
nucleus-nucleus collision. This partonic initial state is regarded
as the quark-gluon matter(QGM) formed in relativistic
nucleus-nucleus collisions. Secondly, the parton rescattering
proceeds. The rescattering among partons in QGM is randomly
considered by the 2$\rightarrow$ 2 LO-pQCD parton-parton cross
sections~\cite{plb70}. In addition, a $K$ factor is introduced here
to account for higher order and non-perturbative corrections.
Thirdly, hadronization happens after parton rescattering. The
partonic matter can be hadronized by the Lund string fragmentation
regime~\cite{jhep} and/or the phenomenological coalescence
model~\cite{cpc183}. Finally, the hadronic matter continues
rescattering until the hadronic freeze-out (the exhaustion of the
hadron-hadron collision pairs). We refer to~\cite{cpc183} for the
details.

With the final state particles provided by the {\footnotesize
PACIAE} model, we can then calculate the production of light nuclei
(anti-nuclei)with the {\footnotesize DCPC} model. In quantum
statistical mechanics~\cite{book}, one cannot precisely define both
position $\vec q\equiv (x,y,z)$ and momentum $\vec p\equiv
(p_x,p_y,p_z)$ of a particle in the six-dimension phase space
because of the uncertainty principle $\Delta\vec q\Delta\vec p
\geqslant h^3 $. We can only say that this particle lies somewhere
within a six-dimension quantum ¡°box¡± or ¡°state¡° with a volume of
$\Delta\vec q\Delta\vec p$. A particle state occupies a volume of
$h^3$ in the six-dimension phase space~\cite{book}. Therefore, one
can estimate the yield of a single particle by defining an integral
$Y_1=\int_{H\leqslant E} \frac{d\vec qd\vec p}{h^3}$, where $H$ and
$E$ are the Hamiltonian and energy of the particle, respectively.
Similarly, the yield of the N particle cluster can be estimated as
the following integral:
\begin{equation}
Y_N=\int ...\int_{H\leqslant E} \frac{d\vec q_1d\vec p_1...d\vec
q_Nd\vec p_N}{h^{3N}}. \label{funct1}
\end{equation}

In addition, equation~(\ref{funct1}) must satisfy the following
constraint conditions:

\begin{equation}
m_0\leqslant m_{inv}\leqslant m_0+\Delta m,
\end{equation}
\begin{equation}
q_{ij}\leqslant D_0,(i\neq j;i,j=1,2,\ldots,N).
\end{equation}
where
\begin{equation}
m_{inv}=\Bigg[\bigg(\sum^{N}_{i=1} E_i \bigg)^2-\bigg(\sum^{N}_{i=1}
\vec p_i \bigg)^2 \Bigg]^{1/2},
\end{equation}
and $E_i$, $\vec p_i$($i$=1,2,\ldots,$N$) are the energies and
momenta of particles, respectively. $m_0$ and $D_0$ stand
for,respectively, the rest mass and diameter of light (anti)nuclei,
$\Delta m$ refers to the allowed mass uncertainty, and $
q_{ij}=|\vec q_{i}-\vec q_{j}|$ is the vector distance between
particles $i$ and $j$. Because the hadron position and momentum
distributions from transport model simulations are discrete, the
integral over continuous distributions in equation~(\ref{funct1})
should be replaced by the sum over discrete distributions.

\section {Results and Discussion}
First we produce the final state particles using the {\footnotesize
PACIAE} model~\cite{cpc183}. In the {\footnotesize PYTHIA}
simulations, we assume that hyperons heavier than $\Lambda$ decayed
already. The model parameters are fixed on the default values given
in {\footnotesize PYTHIA}~\cite{jhep}. However, the $K$ factor as
well as the parameters parj(1), parj(2), and parj(3), which are
relevant to the hadrons production in {\footnotesize PYTHIA}, are
given by fitting the ALICE data of $p$, $\overline p$, $\Lambda$ in
Pb-Pb collisions at $\sqrt{s_{\rm{NN}}}=2.76$~\cite{h1303,prc88m}.
Specific details of this method is similar to the
paper~\cite{prc88chen}. The fitted parameters of $K$= 3(default
value is 1 or 1.5~\cite{jhep}), parj(1) = 0.15(0.1), parj(2) =
0.38(0.3), and parj(3) = 0.65(0.4) are used to generate
$1.833\times10^6 $ minimum-bias events by the {\footnotesize PACIAE}
model for the 0-20\% centrality Pb-Pb collisions at
$\sqrt{s_{\rm{NN}}}=2.76$ TeV with $|y|<0.5$ acceptances, as shown
in Table ~\ref{biao1}.

\begin{table}[htbp]
\caption{The integrated yield dN/dy of particles at midrapidity
($|y|<0.5$), for $p,\bar{p}$ and $\Lambda$ in Pb-Pb collisions of
$\sqrt{s_{\rm{NN}}}=2.76$~TeV with 0-20\% centrality.}
\setlength{\tabcolsep}{19.5pt}
\renewcommand{\arraystretch}{1.4}
\begin{tabular}{ccc} \hline \hline
Particle type  & ALICE$^a$ & PACIAE \\ \hline
$p$  & $25.9\pm1.6$ & $25.9$  \\
$\Bar{ p}$  & $26.0\pm1.8$ & $24.8$ \\
$\Lambda$  & $19.3\pm1.4$ & $19.18$ \\
\hline \hline
\multicolumn{3}{l}{$^a$ The ALICE data are taken from Ref.~\cite{h1303,prc88m}.} \\
\\\end{tabular} \label{biao1}
\end{table}

\begin{table*}[tbp]
\caption{The integrated yields $dN/dy$ of $d$, $\overline d$,
$^3{{He}}$, ${\overline{^3 He}}$, ${_{\Lambda}^3 H}$ and
$\overline{_{\overline\Lambda}^3 H}$ calculated by {\footnotesize
PACIAE+DCPC} model in midrapidity Pb-Pb collisions of
$\sqrt{s_{\rm{NN}}}=2.76$~TeV. $\langle N_{part}\rangle$ is shown
for each centrality bin.}
\setlength{\tabcolsep}{ 14.5pt}
\renewcommand{\arraystretch}{1.4}
\begin{tabular}{ccccccc}
\hline  \hline \cline{1-7} Centrality
 &$0\%-5\%$&$5\%-10\%$&$10\%-20\%$&$20\%-30\%$&$30\%-50\%$&$50\%-80\%$\\ \hline
$\langle N_{part}\rangle$&379&  323&  256&  186&  110&  36 \\ \hline
$d^a$&0.112 & 0.0917  & 0.0703 &  0.0461 & 0.0199 &  0.00345 \\
$\overline{d}^a$&0.101 & 0.0838&  0.0631 & 0.0412 & 0.0182&   0.00311 \\
$^3He^b$ &3.84E-04&  2.95E-04 & 2.13E-04&  1.12E-04 &  4.99E-05&  5.13E-06\\
${\overline{^3 He}^b}$ &3.29E-04 & 2.51E-04 & 1.82E-04 & 9.44E-05 & 4.30E-05 & 4.29E-06 \\
$_{\Lambda}^3 H^c$ &4.47E-05&  3.40E-05 & 2.33E-05 & 1.16E-05 &  5.94E-06 & 5.67E-07 \\
${\overline{_{\overline\Lambda}^3 H}^c}$& 3.24E-05 & 2.45E-05 & 1.71E-05 & 8.26E-06 & 4.38E-06 & 4.03E-07  \\
\hline \hline \multicolumn{7}{l}{$^a$ calculated with $\Delta
m=0.00042$~GeV for
$d$, $\overline{d}$.} \\
\multicolumn{7}{l}{$^b$ calculated with $\Delta m=0.00091$~GeV for
$^3He$, ${\overline{^3 He}}$.} \\
\multicolumn{7}{l}{$^c$ calculated with $\Delta m=0.00040$~GeV for $_{\Lambda}^3H$, ${\overline{_{\overline\Lambda}^3 H}}$.} \\
\\\end{tabular} \label{biao2}
\end{table*}

Then, the integrated yields $dN/dy$ of light (anti)nuclei $d$
($\overline d$), $^3{He}$ ($^3{\overline{He}}$), as well as
${_{\Lambda}^3 H}$ ($\overline{_{\overline\Lambda}^3H}$) are
calculated by the {\footnotesize DCPC} model for each centrality bin
of 0-5\%, 5-10\%, 10-20\%, 20-30\%, 30-50\% as well as 50-80\%, as
shown in Tab.~\ref{biao2}. One can see from Table II that the yields
$dN/dy$ of light (anti)nuclei and (anti)hypertritons calculated by
the {\footnotesize DCPC} model decrease (or increase) with the
increase of centrality(or $N_{\rm {part}}$); the yields of
(anti)nuclei decrease with the increase of mass; and the yields of
anti-nuclei are less than that of its corresponding nuclei.

\begin{figure}[tbp]
\includegraphics[width=0.45\textwidth]{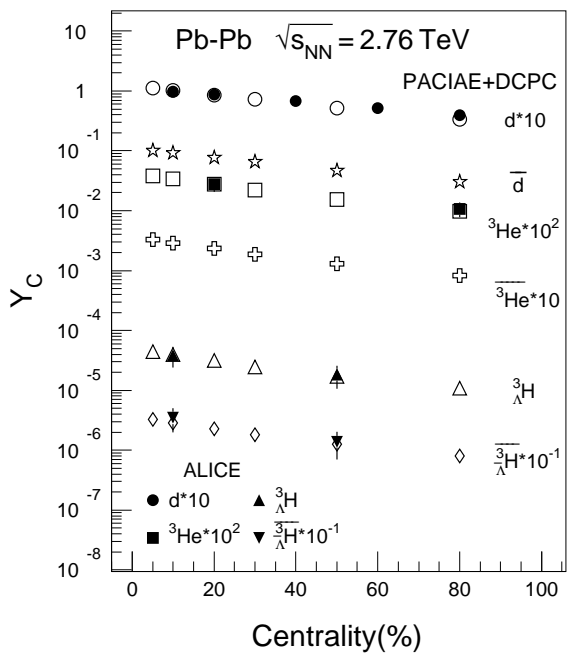}
\caption{The cumulative yields $Y_c$ for light (anti)nuclei and
(anti)hyper-triton as a function of centrality in midrapidity Pb-Pb
collisions at $\sqrt{s_{\rm{NN}}}=2.76$~TeV. The solid symbols are
the experimental data points from ALICE~\cite{08453,08951}.
% and the results of SHM model~\cite{prc88m}.
The open symbols represent the outcome for our {\footnotesize
PACIAE+DCPC} model.} \label{tu1}
\end{figure}

In order to facilitate comparison with the experimental data, the
cumulate yield $Y_c$ is described as
\begin{equation}
Y_c= \frac{1}{C}\int_{0}^{C}\frac{dN}{dy}dc.
\end{equation}
Where $c$ is the value of centrality bins. The Fig.~\ref{tu1}
effectively shows the cumulative yields $Y_c$ of $d$, $\overline d$,
${_{\Lambda}^3 H}$, $\overline{_{\overline\Lambda}^3 H}$, $^3{{He}}$
and $\overline{^3{He}}$ in different centrality bins Pb-Pb
collisions at $\sqrt{s_{\rm{NN}}}=2.76$~TeV. One can see in
Fig.~\ref{tu1}, as well as Tab.~\ref{biao2}, the yields of light
(anti)nuclei and (anti)hyper-triton all decrease rapidly with the
increase of centrality, presenting a fall as the index distribution.
Meanwhile, the {\footnotesize PACIAE+DCPC} model results (the open
symbols) are consistent with the ALICE experimental
data~\cite{08453,08951} (the solid symbols).
%%and the results of SHM model~\cite{prc88m}(the solid symbols).

\begin{figure}[htbp]
\includegraphics[width=0.46\textwidth]{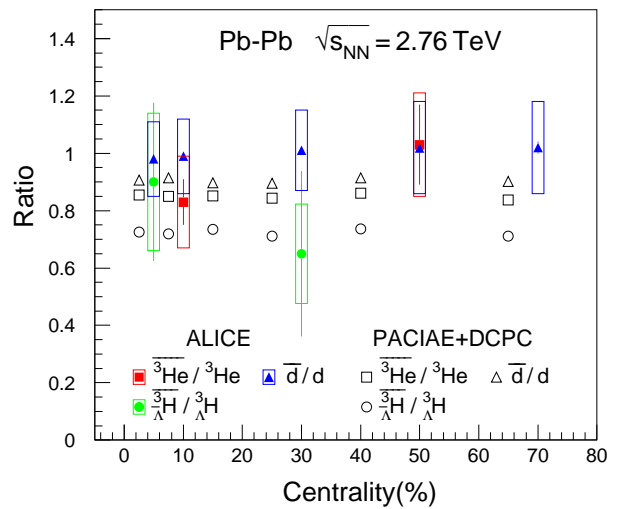}
\caption{Yield ratios of light anti-nuclei and anti-hypertriton
($\overline d$, $^3{\overline{He}}$, and
$\overline{_{\overline\Lambda}^3 H}$) to light nuclei and
hyper-triton ($d$, $^3{{He}}$, and ${_{\Lambda}^3 H}$) in
midrapidity Pb-Pb collisions at $\sqrt{s_{\rm{NN}}}=2.76$~TeV,
plotted as a function of centrality. Open symbols represent our
{\footnotesize PACIAE+DCPC} model results.  Solid symbols are the
experimental data points from ALICE~\cite{08453,08951}, which
statistical uncertainties are represented by bars and systematic
uncertainties are represented by open boxes.} \label{tu2}
\end{figure}

\begin{figure*}[!htbp]
\includegraphics[width=0.90\textwidth]{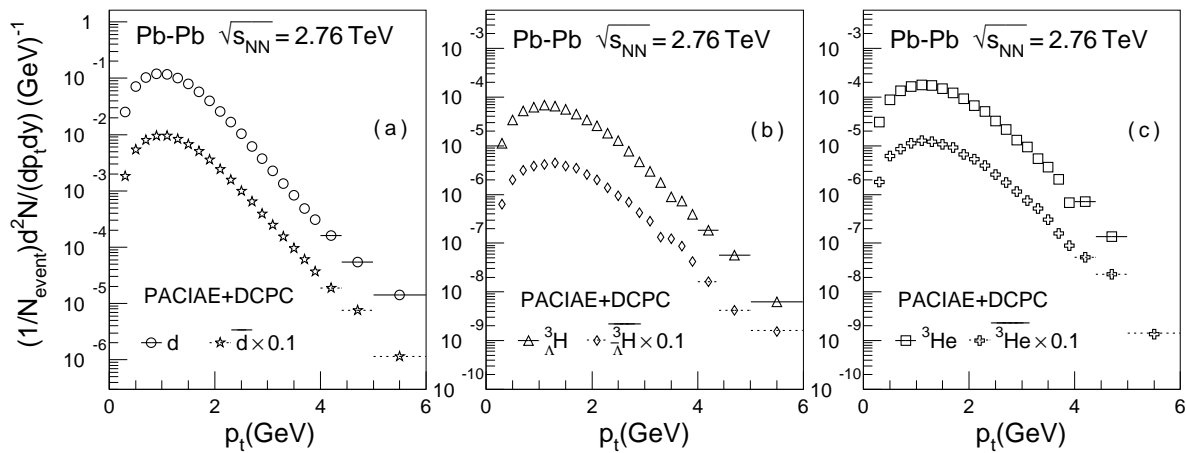}
\caption{The results (open points) of the transverse momentum
distributions of light anti(nuclei) for {\footnotesize PACIAE+DCPC}
model in the 0-10\% most central Pb-Pb collisions at
$\sqrt{s_{\rm{NN}}}=2.76$~TeV with $|y|<0.5$ acceptances, calculated
for (a) $d$ and $\overline d$, (b) $\overline{_{\overline\Lambda}^3
H}$ and ${_{\Lambda}^3 H}$, and (c) $^3{\overline{He}}$ and
$^3{{He}}$.
%% The ALICE data (solid points) are from
%%Ref.~\cite{doc,npa914}.
}\label{tu3}
\end{figure*}

In Fig.~\ref{tu2}, the yield ratios of light anti-nuclei and
anti-hypertriton ($\overline d$, ${\overline{^3He}}$, and
$\overline{_{\overline\Lambda}^3 H}$) to light nuclei and
hyper-triton ($d$, $^3{{He}}$ and ${_{\Lambda}^3 H}$), are given in
different centrality Pb-Pb collisions at
$\sqrt{s_{\rm{NN}}}=2.76$~TeV. To facilitate comparison,
experimental results from ALICE~\cite{08453,08951} are also given
with the solid points. One can see from Fig.~\ref{tu2}, the yield
ratios of light anti-nuclei to light nuclei and anti-hypertriton to
hyper-triton from central to peripheral collisions remain unchanged,
and their corresponding values are respectively about 0.91,
0.85,0.72, although their yields decrease rapidly with the
centrality as shown in Tab.~\ref{biao2} and Fig.~\ref{tu1}. The
results obtained from our model are also in agreement with the
experimental data from ALICE~\cite{08453,08951} within error ranges.
%; although ALICE experimental results has relatively large
%error at low centrality range.

\begin{figure}[htbp]
\includegraphics[width=0.44\textwidth]{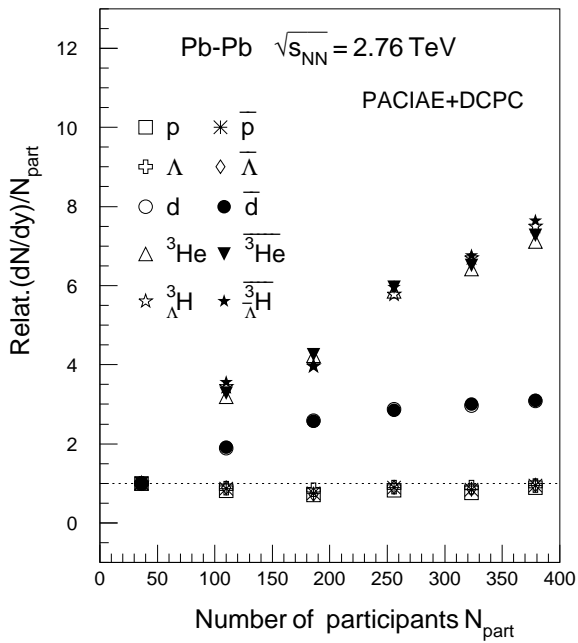}
\caption{ The integrated yield $dN/dy$ as a function of $N_{\rm
part}$. The yields at midrapidity for $p,\overline p,\Lambda,
\overline \Lambda, d, \overline d$, $_{\Lambda}^3H$,
$\overline{_{\overline\Lambda}^3 H}, ^3{He}, ^3{\overline{He}}$ are
divided by $N_{\rm part}$, normalized to the peripheral collisions
(50\%-80\%). The results are calculated by {\footnotesize
PACIAE+DCPC} model in Pb-Pb collisions at
$\sqrt{s_{\rm{NN}}}=2.76$~TeV.} \label{tu4}
\end{figure}

In Fig.~\ref{tu3}, the transverse momentum distributions of $d$,
$\overline d$, ${_{\Lambda}^3 H}$, $\overline{_{\overline\Lambda}^3
H}$, $^3{{He}}$ and ${\overline{^3He}}$ calculated were shown in the
0-10\% central Pb-Pb collisions at $\sqrt{s_{\rm{NN}}}=2.76$ TeV
with $|y|<0.5$ acceptances. Figures 3(a), 3(b) and 3(c) are the
calculated the transverse momentum $p_t$ distributions for
$d$($\overline d$), ${_{\Lambda}^3
H}$($\overline{_{\overline\Lambda}^3 H}$), as well as
$^3{{He}}$($^3{\overline{He}}$), respectively. The pattern of light
(anti)nuclei transverse momentum distributions is similar to the
ALICE~\cite{08951,jpgnpp} of transverse momentum. In addition, you
can see that the the peak of transverse distribution appears at
similar range for nuclei and anti-nuclei situations.
%momentum that   corresponded is nearly equal
%to the one of corresponding anti-nuclei.
%% Quantitatively, the
%%{\footnotesize PACIAE+DCPC} model results(the open symbols) are in
%%agreement with the ALICE data~\cite{doc,npa914}.

\begin{table}[htbp]
\caption{ The results of the light (anti)nuclei average transverse
momentum $\langle p_t \rangle$ calculated by PACIAE+DCPC mode in the
Pb-Pb collisions at $\sqrt{s_{\rm{NN}}}=2.76$ TeV, comparing with
the Au-Au collisions at $\sqrt{s_{\rm{NN}}}=200$ GeV~\cite{prc86}.}
\setlength{\tabcolsep}{4.5pt}
\renewcommand{\arraystretch}{1.4}
\begin{tabular}{ccccccc}
\hline \hline Particle type  & $d$\ \ & $\overline{d}$\ \ & $^3He$\
\ & ${\overline{^3 He}}$ \ \ & $_{\Lambda}^3 H$\ \ &
${\overline{_{\overline\Lambda}^3 H}}$  \\ \hline
Pb-Pb(2.76~TeV)  & $1.00$ & $1.05$ &$1.16$&$1.24$ &$1.17$ &$1.22$     \\
Au-Au(200~GeV)    & $0.92$ & $0.96$ &$1.05$ & $1.06$& $1.15$&$1.18$ \\
\hline \hline
\\\end{tabular} \label{biao3}
\end{table}

Furthermore, the results of light (anti)nuclei average transverse
momentum $\langle p_t \rangle$ calculated by {\footnotesize
PACIAE+DCPC} model in central rapidity Pb-Pb collisions at
$\sqrt{s_{\rm{NN}}}=2.76$ TeV  are given in Table ~\ref{biao3}. As a
comparison, the results~\cite{prc86} of light (anti)nuclei average
transverse momentum $\langle p_t\rangle$ in Au-Au collisions at
$\sqrt{s_{\rm{NN}}}=200$ GeV are also showed. Here we can see that
the average transverse momentum of light anti-nuclei is slightly
larger than that of the corresponding light nuclei. And the results
of light (anti)nuclei average transverse momentum $\langle p_t
\rangle$ in Pb-Pb collisions are larger than one of Au-Au
collisions.

In Fig.~\ref{tu4} we show the integrated yields $dN/dy$ of $d$,
$\overline d$, ${_{\Lambda}^3 H}$, $\overline{_{\overline\Lambda}^3
H}$, ${^3{He}}$ and $\overline{^3{He}}$ divided by $N_{\rm {part}}$,
as a function of $N_{\rm part}$, respectively. All data points are
normalized to the values obtained in the peripheral collisions
(50\%-80\%) with participant numbers 36. We choose the 50\%-80\%
centrality bin as the reference of peripheral collisions because of
the limited statistics of $N_{\rm{part}}$ and yield, as well as the
strong fluctuation in the 80\%-100\% centrality bins. It shows that
the yields of light (anti)nuclei and (anti)hyper-triton per
participant nucleon increase rapidly with the increase of the number
of participant $N_{\rm{part}}$. Obviously, this distribution
properties of light (anti)nuclei and (anti)hyper-triton production
in Pb-Pb collisions at $\sqrt{s_{\rm{NN}}}=2.76$~TeV mainly depend
on their mass number, i.e., the greater the mass number, the faster
the yield increases.

\begin{figure}[htbp]
\includegraphics[width=0.48\textwidth]{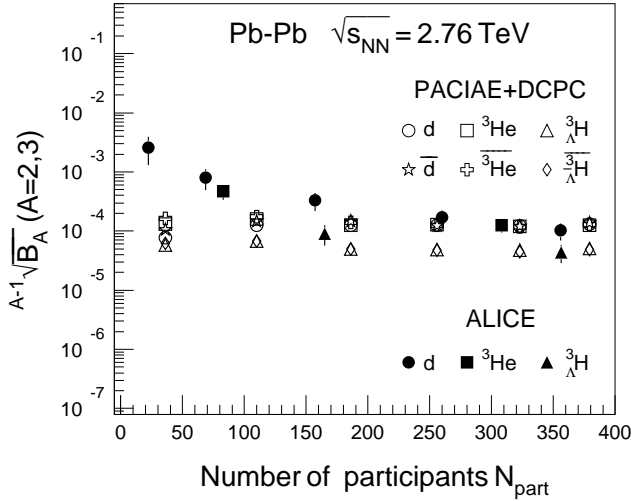}
\caption{Coalescence parameters $B_A$ as a function of $N_{\rm
part}$ for light (anti)nuclei in Pb-Pb collisions at
$\sqrt{s_{\rm{NN}}}=2.76$~TeV. The open symbols represent our mode
results evaluated for different types of (anti)nuclei. The solid
points are the data points from ALICE measurements evaluated with
deuterons, hypertriton and $^3{He}$~\cite{h1303,08453,08951,prc1301}.} \label{tu5}
\end{figure}

In heavy ion collisions, the coalescence process of light
(anti)nuclei, and (anti)hypernuclei is historically described
~\cite{prl1976,prc77,plb79} by the coalescence parameter $B_A$. The
differential invariant yield is related to the primordial yields of
nucleons and is described by the equation:
\begin{equation}
E_A\frac{d^3N_A}{d^3P_A}= B_A(E_P\frac{d^3N_P}{d^3P_P})^A,
\end{equation}
where $Ed^3N/d^3p$ stands for the invariant yields of nucleons or
light (anti)nuclei, and (anti)hypernuclei, and A are the atomic mass
number, respectively. $p_A, p_p$ denote their momentum, with $p_A =
Ap_p$ assumed. The phase-space information of $p(\overline p)$ and
$\Lambda(\overline\Lambda)$, which is used as an input for the
coalescence prescription, can be reproduced by the hydrodynamic
blast-wave model.

The coalescence parameters $B_A$ can be evaluated by comparing the
invariant yields of the light (anti)nuclei and the primordial
(anti)nucleons. Fig. 5 presents the $\sqrt[A-1]{B_A}$ as a function
of $N_{part}$, which remain unchanged from central to peripheral
collisions for our model results. $B_2$ and $\sqrt{B_3}$ calculated
based on the invariant yields of $d$, $\overline d$,
$^3He$($\overline{^3{He}}$), $^3_\Lambda
H$($\overline{_{\overline\Lambda}^3 H}$) are consistent with the
ALICE results~\cite{h1303,08453,08951,prc1301} as $N_{part}\geq
100$. In addition, we noticed that $\sqrt {B_3}$ of $^3_\Lambda
H$($\overline{_{\overline\Lambda}^3 H}$) is smaller than one of
$^3He$($\overline{^3{He}}$) even though the two nuclei have the same
mass number. This reflects the strangeness content dependence of the
coalescence parameter, i.e, there exists an additional penalty
factor due to strangeness.

In a word, we hope utilizing the study of light
(anti)nuclei and light (anti)hyper-triton, like their yields and ratios, their centrality
dependence, their transverse momentum distributions, to seek a
new clue to explore the new production mechanism for heavier
antimatter in relativistic heavy ion collisions.

\section {Conclusion}
In this paper, we use the {\footnotesize PACIAE+DCPC} model to
investigate the properties of light (anti)nuclei and
(anti)hyper-triton production in Pb-Pb collisions at
$\sqrt{s_{\rm{NN}}}=2.76$ TeV with $|y|<0.5$ acceptances. The
integrated yields $dN/dy$ of light (anti)nuclei $d$ ($\overline d$),
$^3{He}$ ($^3{\overline{He}}$), as well as ${_{\Lambda}^3 H}$
($\overline{_{\overline\Lambda}^3H}$) are calculated by the
{\footnotesize DCPC} model for each centrality bin. The results show
that the yields of light (anti)nuclei and (anti)hyper-triton
decrease rapidly with the increase of centrality bins. However, the
yield ratios of light anti-nuclei to light nuclei and
anti-hypertriton to hyper-triton are independent of centrality,
which are irrelevant to the decreasing yields from central to
peripheral collisions. In addition, the transverse momentum
distributions of light anti(nuclei) are given in the 0-10\% most
central Pb-Pb collisions. With respect to light nuclei, the
transverse momentum that the peak of distribution corresponded is
nearly equal to the one of corresponding anti-nuclei. Our model
results are in agreement with the ALICE experimental data. We also
gained that the yields of light (anti)matter per participant nucleon
increase linearly with the increase of the number of participants
with $N_{\rm{part}}$. Obviously, this distribution properties of
light (anti)nuclei and (anti)hyper-triton production mainly depend
on their mass number. At last we calculated and discussed
coalescence parameters $B_A$ of light (anti)nuclei and
(anti)hypernuclei.

\begin{center} {ACKNOWLEDGMENT} \end{center}
Finally, we acknowledge the financial support from NSFC(11475149,
11305144, 11303023). The authors thank Prof. Ben-Hao Sa, PH.D.
Yu-Liang Yan for helpful discussions.

\end{document}